\documentclass{cimento}
\pdfoutput=1
\usepackage{graphicx}

\newcommand{\nuless}{\ensuremath{0\nu\beta\beta}}

\title{Neutrino mass eigenstates and their ordering: a Bayesian approach}
\author{S.~Gariazzo}
\instlist{\inst{}Instituto de F\'isica Corpuscular (CSIC-Universitat de Val\`encia), Paterna (Valencia), Spain}

\begin{document}

\maketitle

\begin{abstract}
\looseness=-1
One of the not-yet determined properties of neutrinos is the ordering of their mass eigenstates.
We combine the available data from neutrino oscillations,
neutrinoless double beta decay
and Cosmic Microwave Background observations
to derive robust constraints on the mass ordering
in a Bayesian context.
Based on \cite{Gariazzo:2018pei}.
\end{abstract}

The most appealing aspect of neutrinos is that they currently represent the only sector
of the Standard Model of Particle Physics
where we can look for new physics,
since we know that they are massive particles from neutrino oscillations but we have no idea
on how their mass is generated.
Among the unknowns about neutrinos,
we can find the ordering of their masses, which can be either
``normal'' (NO), when the lightest neutrino has the largest mixing with the electron flavor neutrino,
or ``inverted'' (IO), when the mixing between the lightest mass eigenstate and the electron neutrino is the smallest.

In the last two years, several studies (see references in \cite{Gariazzo:2018pei})
combined the available neutrino oscillation,
neutrinoless double beta decay (\nuless)
and Cosmic Microwave Background (CMB) observations
to derive constraints on the neutrino mass ordering.
The published results of the combined (Bayesian) analyses show that there is a strong dependence
on the assumptions adopted
when parameterizing the neutrino mass sector (see e.g.\ \cite{Simpson:2017qvj,Caldwell:2017mqu}).
In order to check the robustness of the conclusions and update the previous constraints,
we performed a full Bayesian analysis,
varying the parameterization of the neutrino masses according to the different prescriptions
employed in the existing literature.
We tested two cases for describing the neutrino mass sector:
\textit{(A)} the three mass eigenstates ($m_1$, $m_2$, $m_3$) \cite{Simpson:2017qvj}
or
\textit{(B)} the lightest neutrino mass and the two mass splittings measured by neutrino oscillations
($m_{\rm lightest}$, $\Delta m^2_{21}$, $|\Delta m^2_{31}|$) \cite{Caldwell:2017mqu}.
For each of the two possibilities, we vary also the type of the prior on the neutrino masses,
choosing either a linear or a logarithmic one, and probing that different ranges for the prior
do not alter our conclusions.
The calculation of the Bayesian evidence for the different parameterizations
and their comparison through the Bayes factor (see e.g.\ \cite{Trotta:2008qt})
help us to select the most efficient way to scan the parameter space
given the available experimental data.

The datasets adopted in our analyses consist in the neutrino oscillation data
from the 2017 version of 
\cite{deSalas:2017kay}%
\footnote{The resume of the 2017 status of the analysis,
on which this work is based, can be found in~\cite{globalfit}.
A table with the constraints and some plots are resumed under the ``July 2017'' label.},
\nuless\ constraints as employed in \cite{Caldwell:2017mqu} and
CMB observations from Planck \cite{Adam:2015rua}.

\begin{figure}
\includegraphics[width=\textwidth]{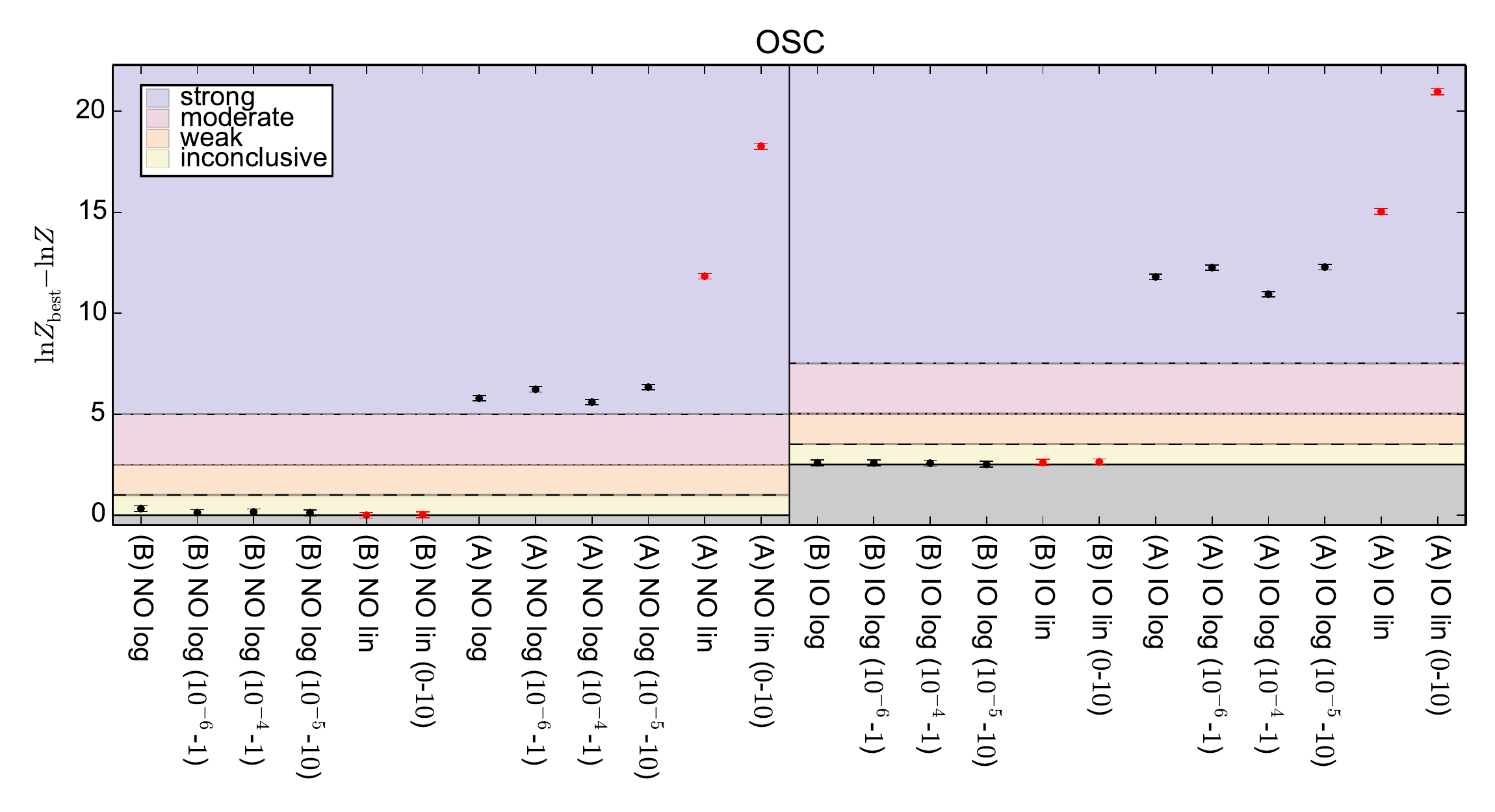}
\caption{\label{fig:evids_osc}
Bayes factors in favor of the preferred model within each panel,
comparing different parameterization and prior choices
for NO (left) and IO (right).
From \cite{Gariazzo:2018pei}.
}
\end{figure}

Figure~\ref{fig:evids_osc} resumes what the neutrino oscillation data alone
tell us about the different parameterizations.
As it is clear from the figure,
choosing the parameterization \textit{A} should be a strongly disfavored choice with respect to using
what we call the case \textit{B}, which is more closely related to the physical parameters measured
by neutrino oscillation experiments (i.e.\ the mass splittings, not the absolute masses).
While the physics of the two cases is the same,
the latter is more efficient when one has to sample the parameter space to study neutrino oscillation data
(there is less waste of parameter space).

\begin{figure}
\includegraphics[width=\textwidth]{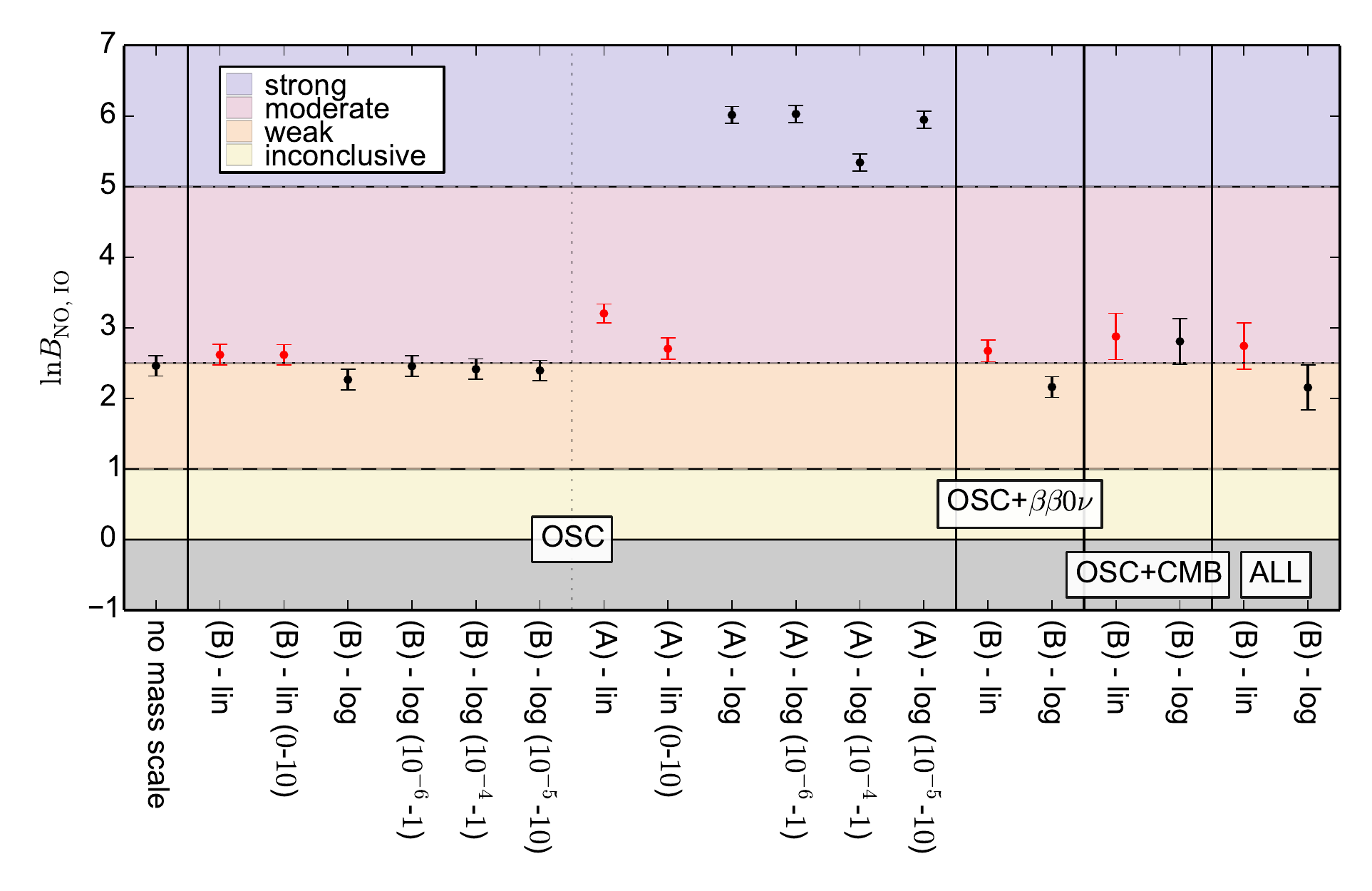}
\caption{\label{fig:bayf}
Bayes factors in favor of the NO scenario, using different parameterizations (cases \textit{A} and \textit{B}),
priors (linear or logarithmic, different ranges)
and data combinations.
From \cite{Gariazzo:2018pei}.
}
\end{figure}

We also used the Bayes factors comparing NO and IO within the same parameterization choices to assess
how much current data prefer normal ordering.
Figure~\ref{fig:bayf} resumes our results, showing that the Bayes factor is rather stable
against variations in the way we explore the neutrino mass sector and the introduction of \nuless\ or CMB data.
Indeed, the weak-to-moderate preference in favor of NO is currently driven by neutrino oscillation data alone,
which were giving a preference for NO at the $2\sigma$ level when we performed the analyses.
There is only one set of calculations for which the results do not agree with what already stated:
when the parameterization \textit{A} and a logarithmic prior are adopted (as in \cite{Simpson:2017qvj}),
the preference for NO becomes strong solely as a consequence of the different allowed parameter space
for the second-to-lightest neutrino mass eigenstate.
This highlights the importance of checking the prior and parameterization choices in order to obtain
reliable and robust results in the context of Bayesian analysis.

A final comment is related to the publication of new experimental measurements
from the T2K and NO$\nu$A experiments at the beginning of 2018.
The updated global analyses of neutrino oscillation data \cite{Capozzi:2018ubv}
(see also the updated \cite{deSalas:2017kay,globalfit}) which take into account these new data, indeed,
show a stronger preference in favor of NO, which is now at the level of $3\sigma$.
A new combined analysis including these new results is presented in \cite{deSalas:2018bym}.

\acknowledgments
Work supported by the European Union's Horizon 2020 research and innovation programme under the Marie Sk{\l}odowska-Curie individual grant agreement No.\ 796941 and by the Spanish grants SEV-2014-0398 and FPA2017-85216-P (AEI/FEDER, UE), and PROMETEOII/2014/084 (Generalitat Valenciana).

\end{document}